\documentclass[aps,prl,twocolumn,showpacs,superscriptaddress,groupedaddress]{revtex4}
\usepackage[normalem]{ulem}
\usepackage{graphicx}  
\usepackage{dcolumn}   
\usepackage{bm}        
\usepackage{amssymb}   
\usepackage[dvipsnames]{xcolor}

\begin{document}

\widetext


\title{Bell States via Two-Particle Contact Interaction: Shannon Entropy as an Indicator of Entanglement Dynamics}

\author{Sandeep Mishra, Anjana Bagga, Anu Venugopalan$^{*}$}

\address{ University School of Basic and Applied Sciences \\
Guru Gobind Singh Indraprastha University, Sector 16 C, Dwarka, New Delhi-110078, India}
\date{\today}

\begin{abstract}

We study the coherent dynamics of two interacting particles in a quantum double-well and show  that the Shannon entropy can be a definitive signature of entanglement as an alternative to concurrence, a connection not reported previously. This physical model, involving tunneling and contact interaction, is akin to the  Hubbard model which explains the physics of interacting particles in periodic potentials. We show that an interplay between tunneling and contact interaction produces Bell like eigenstates ensuring thereby that the concurrence and the Shannon entropy, two quantities with vastly different physical interpretations, develop the same time dependence. Our result, applied to three experimentally realized physical models works for both spatial and spin degrees of freedom and is significant as it provides a novel measure of entanglement applicable to many  systems currently being explored for  scalable quantum computation.

\end{abstract}

\maketitle

The power and potential of quantum information  is rooted in the singularly quantum phenomenon of {\it entanglement}. The concept describes a quantum state of a composite system of two (or more)  particles which is non-separable and  contains correlations that go beyond that predicted by classical theory \cite{schr,epr}. This uniquely quantum possibility has inspired many theoretical and experimental investigations involving the production and coherent manipulation of entangled states   \cite{exp1,exp2,exp3,bloch,mompart,murman} with the aim of exploring new frontiers in computing power \cite{chaung,zel,ekert}.

In this Letter, we propose a novel signature of entanglement through the Shannon entropy, making it an alternative experimental measure. For a two-particle system, Shannon (spatial) entropy captures the spread of the two particle wavefunction (localization) \cite{santos,loc} while  the concurrence  quantifies the degree of entanglement via correlations that reflect whether the two particle state is separable  or not \cite{concurrence}. Thus, the two quantities seem to have vastly different physical interpretations. We examine two-particle entanglement dynamics via strong contact interaction and tunneling in a model akin to the Hubbard model recently explored in several experimental studies \cite{mompart,murman,folling,serwane,tunneling,hubbard} and apply our result to three experimentally realized physical models incorporating spatial and spin degrees of freedom. We show that a measurement of Shannon entropy reflects entanglement dynamics.

To demonstrate our result, we consider a system of two distinguishable particles in a one dimensional infinite square well potential with a $\delta$ barrier at its center. Each particle can tunnel between the two wells. For an opaque barrier \cite{psenn,vugalter} this system is equivalent to two qubits \cite{foot} where $\vert L\rangle$ and $\vert R \rangle$ represent two orthogonal single qubit states, corresponding to left and right localization in the double well. The tunneling eigenstates, $\vert \phi\rangle$ and $\vert \tilde{\phi}  \rangle$ are linear superpositions of $\vert L\rangle$ and $\vert R \rangle$ (see Fig. 1(a)).
The inclusion of contact interaction between the two particles provides a mechanism for entanglement. The system can be described by the Hamiltonian:
\begin{equation}
H=\sum_{i=1,2}  \epsilon_{0} I^{(i)} + \Delta \sum_{i=1,2} \sigma_x^{(i)}+U \sigma_z^{(1)}\otimes \sigma_z^{(2)},
\end{equation}
where $\epsilon_{0} $ is the lowest single particle energy when confined in either well, $\Delta $ is the tunneling strength and $U$ is the  contact interaction strength . The ratio $U/\Delta$ determines the  energy eigenvalue spectrum of the Hamiltonian (1) and the nature of its eigenfunctions (see Fig. 2). For  $U / \Delta \gtrsim 4$, i.e., when the contact interaction dominates the tunneling, the four eigenstates of (1), $\{ \vert ES_i\rangle \}$, become nearly Bell states \cite{chaung} in the two particle  positional basis of $\vert LL \rangle, \vert LR \rangle, \vert RL \rangle $ and $ \vert RR \rangle $ (Fig.1(c)):
\begin{eqnarray}
\vert ES_{1}\rangle, \vert ES_{2}\rangle & \approx & \Psi^{\pm} \approx \frac{1}{\sqrt{2}} (\vert LR\rangle \pm \vert RL\rangle ), \\ \nonumber
\vert ES_{3}\rangle,\vert ES_{4}\rangle & \approx & \Phi^{\mp} \approx \frac{1}{\sqrt{2}}( \vert LL\rangle \mp \vert RR\rangle ).
\end{eqnarray}
The distinctive eigenvalue spectrum of Fig. 2(c) is crucial to the dynamics which correlates the time dependence of the Shannon entropy and concurrence.

In Fig. 2(a), $U=0$ (contact interaction is absent) and hence there is no entanglement and the two particles independently tunnel between wells. The eigenstates here are each equally weighted linear combinations of $\vert LL\rangle,\vert LR\rangle,\vert RL\rangle$ and $\vert RR\rangle$,  with $\vert ES_{2} \rangle $ and $\vert ES_{3} \rangle $ degenerate and separated from $\vert ES_{1} \rangle $ and $\vert ES_{4} \rangle $ by a gap proportional to $ \Delta $. When a repulsive contact interaction is introduced between the two particles, the degeneracy of Fig. 2(a) is lifted and the eigenvalue spectrum  modifies, shifting the states $\vert LL\rangle$ and $\vert RR\rangle$ higher compared to the states $\vert LR \rangle$ and $\vert RL \rangle$ (see Fig. 2(b)). The repulsive interaction makes states $\vert LL \rangle$ and $\vert RR \rangle$ (both particles in the same well), energetically more expensive than $\vert LR \rangle$ and $\vert RL \rangle$ (each particle in a separate well). Consequently, the two lower energy eigenstates have a higher probability amplitude corresponding to $ \vert LR \rangle$ and $\vert RL \rangle$ and the higher energy eigenstates have a higher probability amplitude corresponding to  $\vert LL \rangle$ and $\vert RR \rangle$ as shown in  Fig. 2 where $p > q$ and $p$ and $q$ depend on the parameters $U$ and $\Delta$. When the contact interaction is much stronger than the tunneling strength ($U/\Delta \gtrsim 4$, see Fig. 2(c)) the eigenstates  become nearly  Bell states (2), with the Bell states $\vert \Phi^{\pm}\rangle$ (in $|LL \rangle$ and $|RR \rangle$) widely separated from  $\vert \Psi^{\pm}\rangle$ (in $|LR\rangle$ and $|RL\rangle$) by an amount proportional to $2U$. Additionally, $ E_{4}-E_{3} \approx E_{2}-E_{1} \ll U$ as illustrated in Fig. 2(c). The ratio $U/\Delta \gtrsim 4$ is close to the parameter regimes explored in several recent experiments \cite{murman,folling,shinkai,oliveira,spin}.

We now show how the unique eigenvalue spectrum of Fig. 2(c) is instrumental in making the Shannon (spatial) entropy mimic entanglement dynamics. Consider a general two-particle  initial state
\begin{equation}
\psi(x_{1},x_{2},0)=\sum_{i=1,4} c_{i}\vert ES_{i} \rangle,
\end{equation}
where $x_1$ and $x_2$ are the position coordinates of each particle and $\lbrace c_{i}\rbrace s$ are the overlap amplitudes of the initial state with the eigenstates $\{\vert ES_{i} \rangle\} $. The time evolution (3), driven by the eigenvalue spectrum of (1) is:
\begin{equation}
\psi(x_{1},x_{2},t)=\sum_{i=1,4} c_{i}\vert ES_{i} \rangle e^{-iE_{i}t/\hbar},
\end{equation}
where $\{E_i\}$s are energy eigenvalues and $t$ is the time. Eq. (4), expressed in the positional basis of $\vert LL\rangle,\vert LR\rangle,\vert RL\rangle$ and $\vert RR\rangle$ is:
\begin{eqnarray}
\psi(x_{1},x_{2},t)&=& \alpha_{LL}\vert LL \rangle + \alpha_{LR}\vert LR \rangle \\ \nonumber
&& + \alpha_{RL}\vert RL \rangle + \alpha_{RR}\vert RR \rangle,
\end{eqnarray}
where $\alpha_{LL}=\sum_{i=1,4} c_{i}\langle LL \vert ES_i\rangle e^{-iE_{i}t/\hbar}$ represents the overlap amplitude of the state $|LL\rangle$ with the eigenstates $\{|ES_i \rangle\}$ and $\alpha_{LR}$, $\alpha_{RL}$ and $\alpha_{RR}$ can each be defined in a similar manner. When $U / \Delta \gtrsim 4$, the eigenstates $\{|ES_{i} \rangle\}$ become nearly Bell states (Fig. 2(c)) with $\alpha_{LL}$ and $\alpha_{RR}$ such that $c_1 \approx 0$, $ c_2 \approx 0$, $c_3 \neq 0$, $c_4 \neq 0$, while for $\alpha_{LR}$ and $\alpha_{RL}$, $c_1 \neq 0$, $c_2 \neq 0$, $c_3 \approx 0$, $c_4 \approx 0$.
From the overlap amplitudes $\alpha_{LR}$, $\alpha_{RL}$ , $\alpha_{LL}$, and $\alpha_{RR}$ the corresponding probabilities, $P_{LR}$, $P_{RL}$,  $P_{LL}$ and $P_{RR}$ for the state (5) are :
\begin{eqnarray}
P_{LR}, P_{RL}=\frac{c_{1}^{2}}{2}+\frac{c_{2}^{2}}{2}\pm c_{1}c_{2} \cos\left[\frac{(E_{2}-E_{1})t}{\hbar}\right], \\ \nonumber
P_{LL}, P_{RR}=\frac{c_{3}^{2}}{2}+\frac{c_{4}^{2}}{2} \pm c_{3}c_{4} \cos\left[\frac{(E_{4}-E_{3})t}{\hbar}\right].
\end{eqnarray}
It is clear from (6) that the probabilities vary sinusoidal with time, governed only by the low frequencies shown in Fig. 2(c).  $P_{LR},P_{RL},P_{LL}$ and $P_{RR}$  capture probabilities for the presence (spatial localization) of the two particles in the right and left wells as a function of time and determine the two-particle spatial (Shannon) entropy:
\small
\begin{eqnarray}
S_{H}(t) &=& - \{ P_{LL}\log_{2}P_{LL}+P_{LR}\log_{2}P_{LR}  \nonumber \\
&& + P_{RL}\log_{2}P_{RL}+P_{RR}\log_{2}P_{RR} \}
\end{eqnarray}
\begin{eqnarray}
\approx A \left\{ \frac{c_{3}^{2}}{2}\frac{c_{4}^{2}}{2} \cos\left[\frac{2(E_{4}-E_{3})t}{\hbar}\right]
+ \frac{c_{1}^{2}}{2}\frac{c_{2}^{2}}{2} \cos\left[\frac{2(E_{2}-E_{1})t}{\hbar}\right ] \right\}  + C_{0}, \nonumber
\end{eqnarray}
\normalsize
where $A$ and $C_0$ are constants. Spatial (Shannon) entropy as understood above measures localization in the state space  (defined here by $\vert L \rangle$ and $\vert R \rangle$ for each particle) of the two-particle system and is zero if the two particles are both localized in either well, i.e., if only one of the probabilities in (7) is non zero. Also, $S_{H}$ is maximum when it is equally probable for each particle to be in both wells. We emphasize that the time dependence of the spatial (Shannon) entropy as seen in (7) is governed only by low frequencies shown in  Fig. 2(c).

Concurrence \cite{concurrence}, a standard measure for quantifying entanglement in pure bipartite systems can be written for the two-particle state (5) as:
\begin{equation}
C(\psi)= 2 | \alpha_{LL} \alpha_{RR}-\alpha_{LR} \alpha_{RL}|,
\end{equation}
with $C>0$ corresponding to entangled states and  $C=1$ corresponding to maximally entangled states (Bell states). The modulus square of the concurrence is
\small
\begin{eqnarray}
|C|^{2} = \left \{\frac{c_{3}^{2}}{2}\frac{c_{4}^{2}}{2} \cos\left [\frac{2(E_{4}-E_{3})t}{\hbar}\right] + \frac{c_{2}^{2}}{2}\frac{c_{1}^{2}}{2} \cos\left [\frac{2(E_{2}-E_{1})t}{\hbar}\right] \right \}_{\bf L} \nonumber \\
+ \bigg\{ \frac{c_{2}^{2}}{2}\frac{c_{4}^{2}}{2} \cos \left[\frac{2(E_{4}-E_{2})t}{\hbar}\right] + \frac{c_{3}^{2}}{2}\frac{c_{2}^{2}}{2} \cos\left[\frac{2(E_{3}-E_{2})t}{\hbar}\right]  \\
+ \frac{c_{1}^{2}}{2}\frac{c_{4}^{2}}{2} \cos\left [\frac{2(E_{4}-E_{1})t}{\hbar}\right] + \frac{c_{3}^{2}}{2}\frac{c_{1}^{2}}{2} \cos\left[\frac{2(E_{3}-E_{1})t}{\hbar}\right] \bigg\}_{\bf H} + C_{1}, \nonumber
\end{eqnarray}
\normalsize
where $C_{1}$ is a constant. Eq. (9) is a sum of terms containing all transition frequencies seen in the energy spectrum emerging when $U/\Delta \gtrsim 4$ (Fig. 2(c)). This allows us to put the terms in  $\vert C \vert ^{2}$  into two categories: low $(\textbf{L})$ and high (\textbf{H}) frequency terms. Further, it is possible to write the high frequency part (\textbf{H}) as  sums of cosine terms whose arguments differ by small amount $\Delta \omega ( \approx (E_{4}-E_{3})/ \hbar \approx (E_{2}-E_{1})/ \hbar)$, as is usually seen in the beats phenomenon. This way, we can show that these terms combine such that $ |C|^{2}$ has a form that resembles a beat-like phenomenon:
\begin{eqnarray}
\vert C \vert^{2} &=& B \bigg\{ \frac{c_{3}^{2}}{2}\frac{c_{4}^{2}}{2} \cos \left[ \frac{2(E_{4}-E_{3})t}{\hbar} \right] \nonumber \\
 && + \frac{c_{2}^{2}}{2}\frac{c_{1}^{2}}{2} \cos \left[ \frac{2(E_{2}-E_{1})t}{\hbar} \right] + \alpha \bigg\} _{\bf I} \nonumber \\
&& \times {\left\{ \beta \cos\left[\frac{2(E_{3}-E_{1})t}{\hbar} + \theta \right] + 1 \right \}}_{\bf II}
\end{eqnarray}
where $ B$, $\alpha$ and $\beta$ are constants and $ \beta =  \left \{ (c_{3}^{2}+c_{4}^{2})(c_{1}^{2}+c_{2}^{2})  -  \vert c_{4}^{2}-c_{3}^{2}\vert \vert c_{2}^{2}-c_{1}^{2} \vert \right \}/ \left \{2\left( c_{3}^{2}c_{4}^{2} + c_{1}^{2}c_{2}^{2} \right)\right\}$ and $\theta$ is a function of $\{c_i\}s$ and time.
The expression for $\vert C \vert ^{2}$, (10), clearly contains a high frequency term ({\bf II}) modulated by a slowly varying envelope term ({\bf I}) (see Figs. 3 and 4).
Eqs. (7) and (10) capture the time dependence of the Shannon entropy and the square of the concurrence, respectively, for (1) when $U / \Delta \gtrsim 4$. The envelope in (10) captures amplitude variations  of the value of the square of the concurrence. More significantly, (7) and (10)  show that the time dependence of the envelope of the square of the concurrence (and hence of the concurrence) is similar to that of the spatial (Shannon) entropy, implying thereby that an experimental measurement of spatial entropy can clearly be an alternative indicator of entanglement. Next, we apply this result to three different experimentally realized physical models.

First, we look at a system of two neutral cold atoms trapped in an optically created double well, explored in many recent  experiments \cite{mompart,tunneling,folling,serwane,murman}. Advanced techniques allow the control of interparticle interactions between the atoms with high precision \cite{folling,serwane,murman}. Recently, Murmann et al. have trapped two ultracold Lithium atoms (each in two different hyper-fine states, $ \vert \uparrow\rangle = \vert F=\frac{1}{2} m= +\frac{1}{2}\rangle$ and  $ \vert \downarrow \rangle = \vert F=\frac{1}{2} m= -\frac{1}{2}\rangle $) into the motional ground state of an optical dipole trap mimicing a double well and introduced a repulsive interparticle interaction between them \cite{murman}. They independently control this interparticle interaction strength as well as  the tunneling rate between the two wells \cite{murman}. 

To compare the dynamics of the spatial entropy and the envelope of square of the concurrence, the values of the parameters $\Delta$ (tunneling strength) and $U$(contact interaction strength) used in the experiment of Murmann et al are chosen for the Hamiltonian (1)(see Table I). Fig. 3 shows plots of the spatial (Shannon) entropy and square of the concurrence for four different initial states. In Fig. 3(a) the initial state is $\vert LL\rangle $ for which the coefficients, $\{c_i\}s$ are such that $ c_{1}= 0$,  $c_{2} = 0$ and $ c_{3}= c_{4}=1/\sqrt{2} $. Thus, the time evolution of $\vert LL\rangle $ is governed only by the eigenstates $\vert ES_{3}\rangle$ and $\vert ES_{4} \rangle$ involving only the low frequency $(E_{4}-E_{3})/\hbar$ (see Fig. 2(c)). As a result both  spatial entropy and concurrence have the same sinusoidal time dependence, governed by $2(E_{4}-E_{3})/\hbar$. This is also evident when the values of the coefficients, $\{c_i\}s$, are put in Eqs. (7) and (10). The absence of high frequency oscillations in the square of the concurrence for the initial state $\vert LL\rangle $  can be appreciated by noting that in Eq.(8), $\alpha_{LR}$ and $\alpha_{RL}$ are zero for $\vert LL\rangle $ and hence there is no interference between the two terms.

Fig. 3(b) shows  spatial (Shannon) entropy and square of the concurrence for an initial state $\frac{1}{\sqrt{2}} \lbrace \vert LL\rangle + \vert LR\rangle \rbrace $ with $ c_{1}=c_{2}=c_{3}=c_{4}=1/2 $. Here, the square of the concurrence shows fast oscillations modulated by a slow envelope whose time dependence is similar to that of the spatial entropy. The time dependence of the spatial entropy and the envelope of the square of the concurrence, as seen from (7) and (10), are determined  by a summation of two cosine terms, where the argument of cosine terms are small and nearly equal (i.e., $E_{4}-E_{3} \approx E_{2}-E_{1}$ (see Fig. 2(c)). The fast oscillations in the square of the concurrence is due to the high frequency term in Eq.(10) which  arises due to the interference between two terms in Eq.(8). Fig. 3(c) shows the plots for the initial state $\frac{1}{\sqrt{3}} \lbrace \vert LL\rangle + \vert RR\rangle +  \vert LR\rangle \rbrace $ with $ c_{1}= c_{2}=1/\sqrt{6}, c_{3}=0,c_{4}=\sqrt{2/3}$ where it can again be seen that spatial entropy and envelope of square of the concurrence have similar time dependence for the same reasons as Fig. 3(b). Fig. 3(d) shows the plots for the initial state $\frac{1}{2} \lbrace \vert LL\rangle +  \vert RR\rangle+  \vert LR\rangle +  \vert RL \rangle \rbrace $ with $ c_{1}= 1/\sqrt{2}, c_{2}=0, c_{3}=0 ,c_{4}=1/\sqrt{2} $. Here both spatial entropy and the envelope of the square of the concurrence are constant with time which can be verified by  putting the values of the coefficients, $\{c_i\}s$ in (7) and (10). Alternately, one can appreciate this by observing that the time evolution of the initial state here is ${1/2(\vert LR \rangle + \vert RL \rangle)}e^{−iE_{1}t/\hbar} + {1/2(\vert LL \rangle + \vert RR \rangle)}e^{−iE_{4}t/\hbar} $, where all the probabilities $ P_{LL}, P_{RR}, P_{LR}, P_{RL} $  become independent of time, making the spatial entropy and the envelope of the concurrence independent of time. The high frequency oscillations continue to be present in the concurrence here as the two interfering terms in (8) are non zero. Thus, we conclude that for any initial state the time dependence of the spatial (Shannon) entropy and the envelope of concurrence is similar and they attain their maxima and minima at the same times.

Next, we apply our result to a physical system simulating quantum magnets. Laser cooling techniques that trap cold atoms into optical lattices allow for the study of not only motional states of atoms but also internal spin degrees \cite{spin,spinnew}. Recently, Friedenauer et al have  explored quantum magnetization for two confined  spin-1/2 particles  where entanglement dynamics arises from an interplay between tunneling and spin-spin interaction \cite{spin}. In their system, each spin is represented by two hyperfine ground levels of trapped $^{25} Mg^{+}$ ions in the $^{2}S_{1/2}$ state. The spin states  $ \vert \downarrow\rangle = \vert F=3; m_{f}= 3\rangle$ and  $ \vert \uparrow \rangle = \vert F=2; m_{f}= 2\rangle $ are analogous to $|L \rangle$ and $|R \rangle$ in  our previous example of cold atoms, and $ \frac{1}{\sqrt{2}} \lbrace \vert \uparrow \rangle  \pm \vert \downarrow \rangle \rbrace$ are analogous to the tunneling eigenstates in position, $ \frac{1}{\sqrt{2}} \lbrace \vert L \rangle  \pm \vert R \rangle \rbrace$, resulting from the interaction of each  spin  with a uniform magnetic field  $B_{x}$ \cite{spin}. $B_{x}$ is simulated by coherently coupling the hyperfine levels via laser and radio frequency radiation and is analogous to the tunneling strength, $ \Delta$.  The spin-spin interaction, $U$, is experimentally realized  by interacting the $^{25}Mg^{+}$ ions  with laser beams which induce a state dependent optical dipole force via the ac stark shift \cite{spin}. It may be noted that the Shannon entropy (7) here derives from $P_{\uparrow \uparrow}$ $P_{\uparrow \downarrow}$, $P_{\downarrow \uparrow}$, $P_{\downarrow \downarrow}$, corresponding to spin. The actual experimental parameters $B_{x}$ and $U$ (see Table I) are used to compare concurrence and Shannon (spin) entropy and we find that they show a similar time dependence. This holds true for any initial state and is illustrated in Fig. 4(b) for the initial state $\frac{1}{\sqrt{2}} \left\{\vert \uparrow\uparrow \rangle + \vert \downarrow\downarrow \rangle\right\}$. 

Finally, we demonstrate our results for two electrons confined in quantum molecules, which are double quantum dot (DQD) structures where an electron can tunnel from one quantum dot to another \cite{hayashi,petta,shinkaiold,shinkai}. Such systems are  promising candidates for quantum hardware in quantum computing. Shinkai et al \cite{shinkai} have recently investigated the dynamics of two electrons confined in separate quantum molecules mimicking the model described in this work. The two electrons can each tunnel between the two wells of a DQD and can also interact electrostatically with each other \cite{shinkai}. Both the tunneling  strength, $\Delta$, and the electrostatic interaction strength, $U$, can be experimentally controlled. Oliveira and Sanz have recently analyzed the  two particle entanglement dynamics for this experiment \cite{oliveira}. As before, an interplay of tunneling and interaction leads to the formation of Bell states in the  basis $\vert LL\rangle$, $\vert LR\rangle$, $\vert RL\rangle$ and $\vert RR \rangle$. We use the  parameter values for  $\Delta$  and $U$ from the experiment of Shinkai et al (see Table I) to show that the spatial entropy and the envelope of the concurrence have the same time dependence and attain their maxima and minima at the same times (see Fig. 4(c)).

To conclude, our study of the coherent dynamics of two particles in a quantum double-well shows that the concurrence, a measure of entanglement, can be written as a beat like phenomenon with fast oscillations modulated by a slowly varying envelope when the contact interaction dominates the tunneling strength. More significantly, we show that the Shannon entropy, conventionally reflecting the spread in probabilities of the two particle state, has the same time dependence as that of the envelope of the concurrence, making it a definitive measurable signature of entanglement. This connection between the concurrence and the Shannon entropy, two quantities with vastly different physical interpretations, has not been reported previously. Our model is akin to the Hubbard model that explains the  physics of interacting particles in periodic potentials and ultracold atoms in optical traps simulating qubits and quantum gates. We illustrate our result in three experimentally realized physical models having both spatial and spin degrees of freedom. Our result gives a novel, experimentally accessible measure of entanglement applicable in physical systems that can be explored for quantum information and scalable quantum computation.
\\

$^{*}$Corresponding author\\
anu.venugopalan@gmail.com

\begin{figure}[p]
\includegraphics[width= 8.6 cm]{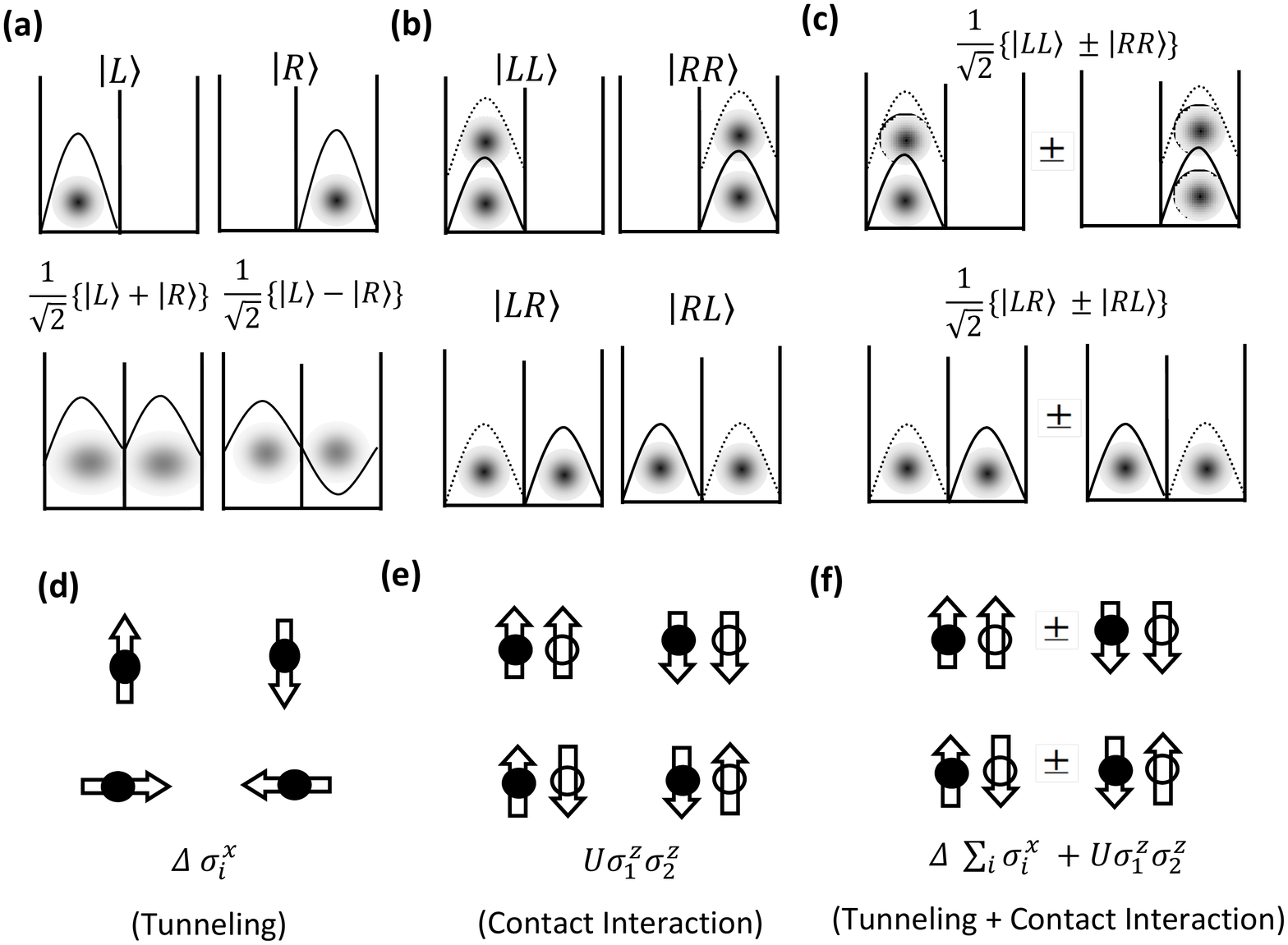}
\caption{\label{fig:epsart1} (a) Infinite square well with $ \delta $ barrier in the center; single particle states $\vert L\rangle$ and $\vert R \rangle$; tunneling eigenstates states $\frac{1}{\sqrt{2}}\lbrace \vert L\rangle \pm \vert R\rangle \rbrace $. (b) Two particle eigenstates $\vert LL \rangle,\vert LR \rangle,\vert RL \rangle $ and  $\vert RR \rangle $ in the presence of contact interaction. (c) Eigenstates of $ \Delta \sum_{i=1,2} \sigma_x^{(i)}+U \sigma_z^{(1)}\otimes \sigma_z^{(2)} $ for $U / \Delta \gtrsim 4 $. (d), (e), (f) corresponding states for spin.}
\end{figure}

\begin{figure}[p]
\includegraphics[width= 8.6 cm]{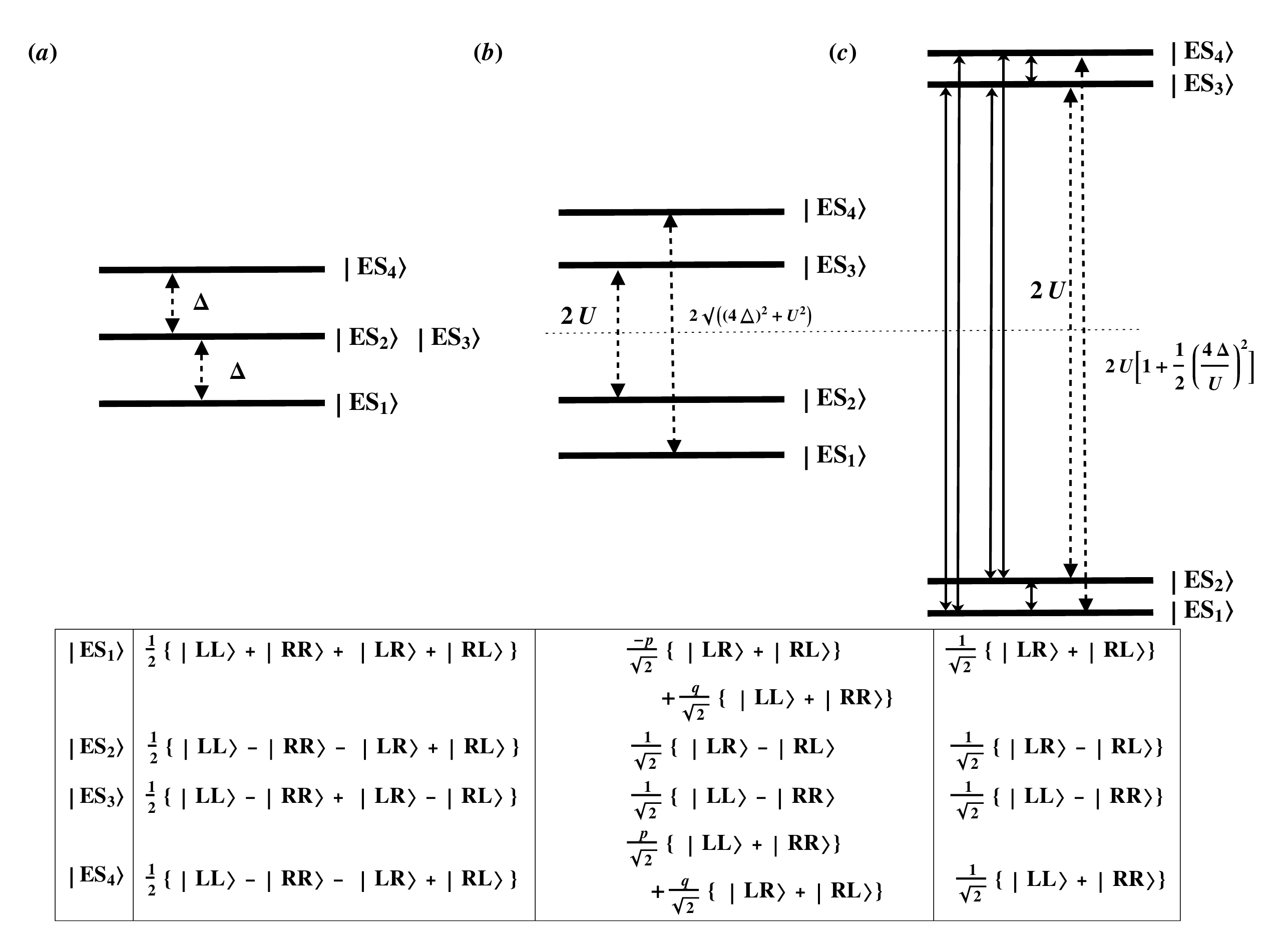}
\caption{\label{fig:epsart2} Energy eigenvalues and eigenstates for (a) $U=0$, eigenstates  equally weighted superpositions of $\vert LL \rangle, \vert LR \rangle, \vert RL \rangle $ and $ \vert RR \rangle $ (b) $ U \approx \Delta $, $\vert LL\rangle$ and $\vert RR\rangle$ raised in energy compared to $\vert LR\rangle$ and $\vert RL\rangle$ due to contact interaction making the probability amplitudes for $\vert LR\rangle$ and $\vert RL\rangle$ higher than $\vert LL\rangle$ and $\vert RR\rangle$ in $\vert ES_{1} \rangle $, $\vert ES_{2}\rangle$ and vice versa for $\vert ES_{3}\rangle$, $\vert ES_{4}\rangle $ $(p > q)$, (c) $ U / \Delta \gtrsim 4 $, eigenstates become nearly Bell states and $\vert ES_{1} \rangle $, $\vert ES_{2} \rangle$ are widely separated from $\vert ES_{3}\rangle$, $\vert ES_{4}\rangle$ by an amount proportional to $2U$.}
\end{figure}

\begin{figure}[p]
\includegraphics[width= 8.6 cm]{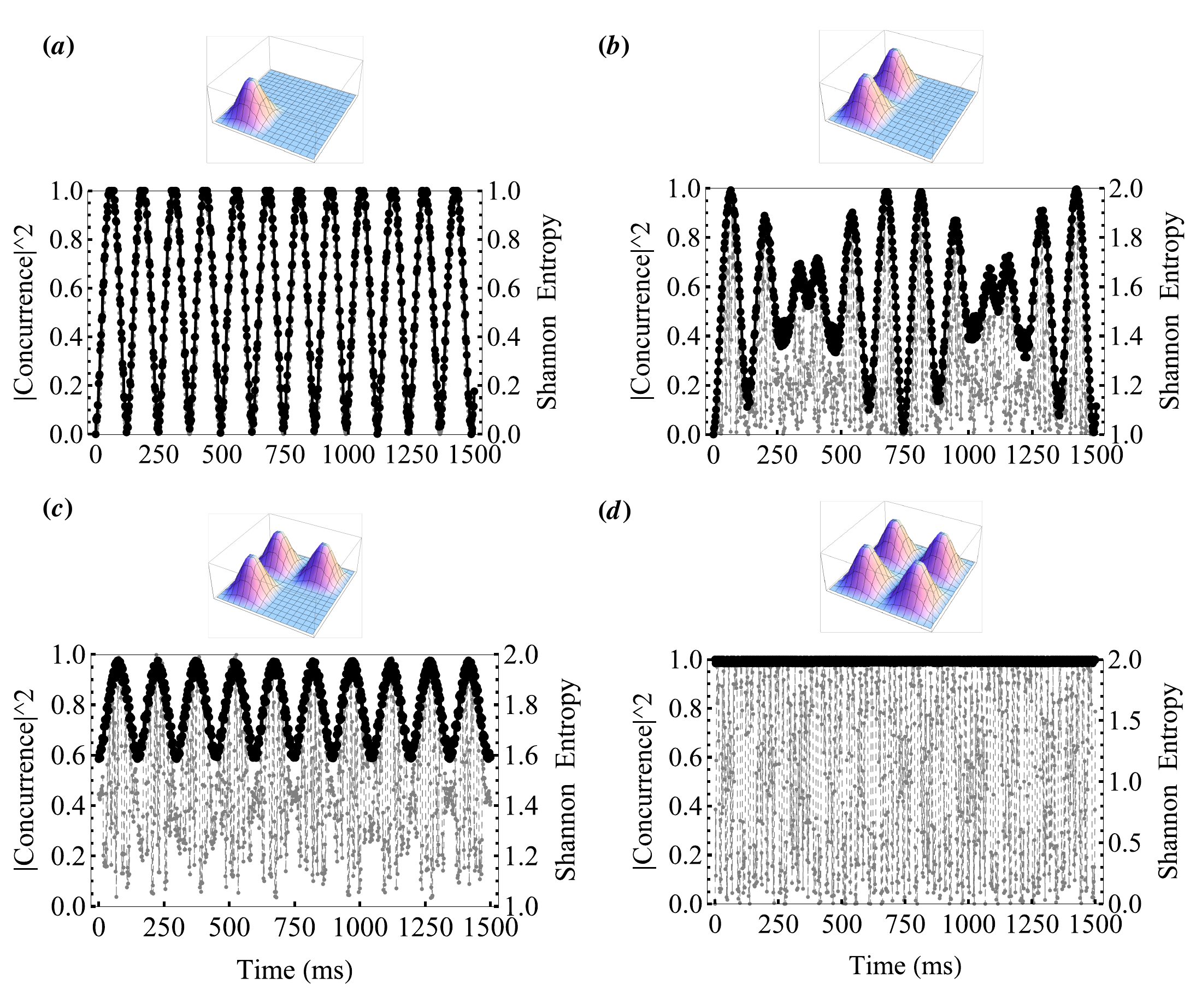}
\caption{\label{fig:epsart} Time evolution of Shannon (spatial) entropy (black) and $\vert$concurrence$\vert ^{2}$ (gray) for four initial states (a) $\vert LL\rangle $ (b) $\frac{1}{\sqrt{2}} \lbrace \vert LL\rangle + \vert LR\rangle \rbrace $ (c) $\frac{1}{\sqrt{3}} \lbrace \vert LL\rangle + \vert RR\rangle +  \vert LR\rangle \rbrace$ (d) $\frac{1}{2} \lbrace \vert LL\rangle +  \vert RR\rangle+  \vert LR\rangle +  \vert RL \rangle \rbrace $. The envelope of $\vert$concurrence$\vert ^{2}$ has a time dependence similar to the Shannon (spatial) entropy.}
\end{figure}

\begin{figure}[p]
\includegraphics[width= 8.6 cm]{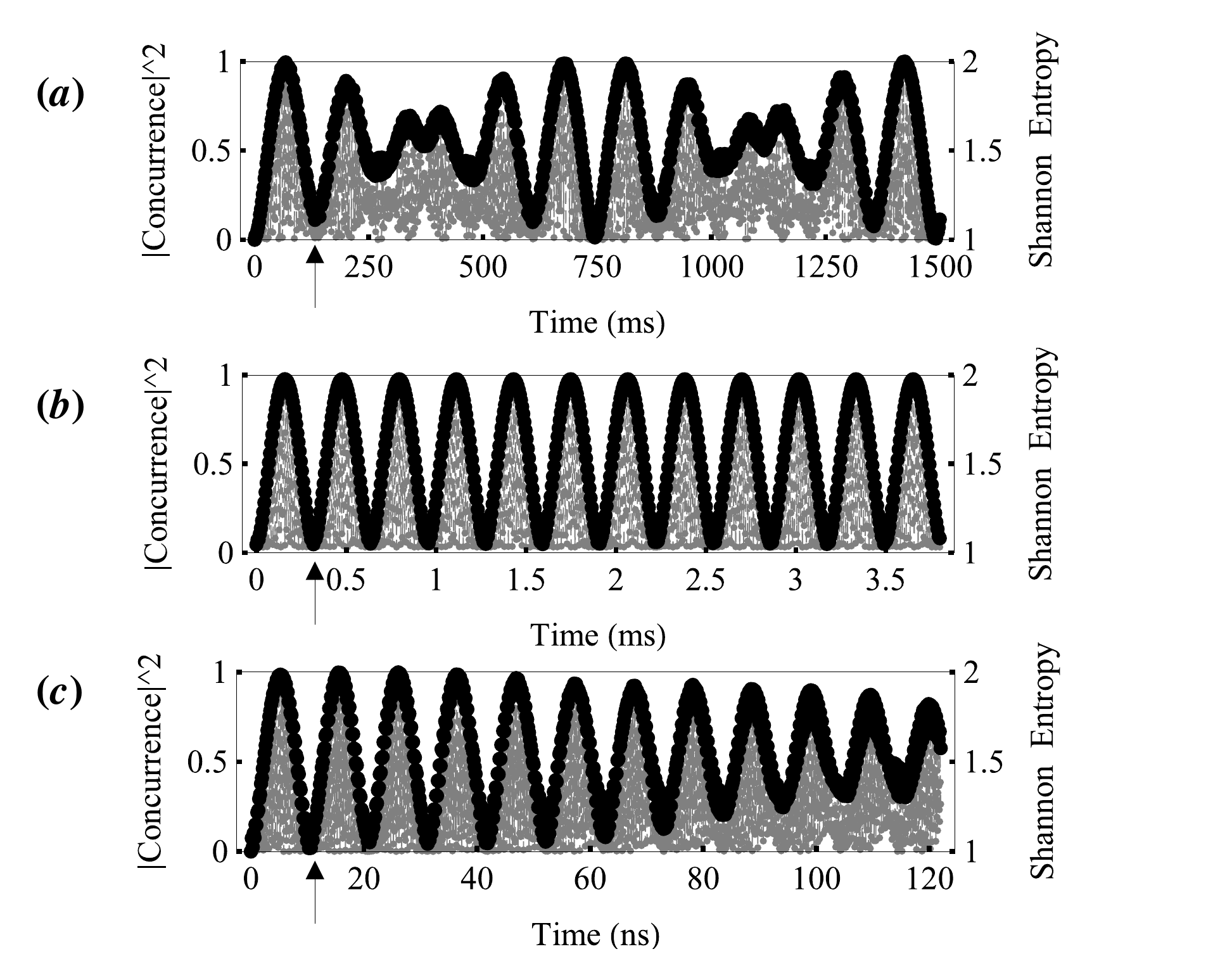}
\caption{\label{fig:wide} Time evolution of Shannon entropy (black) and $\vert$concurrence$\vert ^{2}$ (gray) for the initial state $\frac{1}{\sqrt{2}} \left\{\vert LL\rangle + \vert RR\rangle \right\}$ $(\frac{1}{\sqrt{2}} \left\{\vert \uparrow\uparrow \rangle + \vert \downarrow\downarrow \rangle\right\})$ for (a) Two $^{6}Li$ atoms in an optically created double well,  (b) Two spins (two $^{25}Mg^{+}$ ions in different hyperfine ground states), (c) Two electrons in two separate DQDs. The envelope seen for each system is given by Eq. (7) and is a sum of cosines with frequencies $\omega_a=\frac{E_4-E_3}{\hbar}$ and  $\omega_b=\frac{E_2-E_1}{\hbar}$, the values of $\omega_a$ and $\omega_b$ specific to each system. The Shannon entropy and the envelope of the $\vert$concurrence$\vert ^{2}$ have a similar time dependence. The arrows indicate the time scales $(\frac{h}{2(E_{4}-E_{3})}) $.} 
\end{figure}

\begin{table*}[p]
\caption{\label{tab:table1}Details and parameter values for three physical models}
\begin{ruledtabular}
\begin{tabular}{ccccccc}
System & Particles  & Bell state basis & $U$(eV) & $\Delta$ (eV) & $U/\Delta$ &  Time Scale $(\frac{h}{2(E_{4}-E_{3})}) $(s)\\ \hline
 Optical trap & Neutral atoms $(^{6}Li)$  & Positional & $2.7 \times 10^{-12}$ & $2.66\times 10^{-13}$ & $10$ &  124 ms \\
 Quantum magnet & Ion$(^{25}Mg^{+})$  & Spin & $ 91.1 \times 10^{-12}$ & $ 17.5 \times 10^{-12} $ & $5.2$ &  0.31 ms \\
 Semiconductor DQD & Charged particle $(e^{-})$ & Positional & $25\times 10^{-6}$ & $6.25 \times 10^{-6}$ & 4 &  10.1 ns\\
\end{tabular}
\end{ruledtabular}
\end{table*}

\end{document}